\documentclass[10pt,english,aps,p9,notitlepage,nofootinbib]{revtex4-1}
\usepackage{amsmath}
\usepackage{esint}
\usepackage{graphicx}
\usepackage{hyperref}

\newcommand{\ave}[1]{\left\langle #1 \right\rangle}
\newcommand{\GeV}{\,\mathrm{GeV}}

\begin{document}


\preprint{This line only printed with preprint option}

\title{Multiplicity dependent and non-binomial efficiency corrections
for particle number cumulants}

\author{Adam Bzdak}
\email{bzdak@fis.agh.edu.pl}
\affiliation{AGH University of Science and Technology\\
Faculty of Physics and Applied Computer Science\\ 
30-059 Krak\'ow, Poland}

\author{Romain Holzmann}
\email{r.holzmann@gsi.de}
\affiliation{GSI Helmholtzzentrum f\"ur Schwerionenforschung GmbH,\\ 
64291 Darmstadt, Germany}

\author{Volker Koch}
\email{vkoch@lbl.gov}
\affiliation{Nuclear Science Division\\
Lawrence Berkeley National Laboratory\\
Berkeley, CA, 94720, USA}

\begin{abstract}
In this note we extend previous work on efficiency corrections for
cumulant measurements \cite{Bzdak:2012ab,Bzdak:2013pha}. We will
discuss the limitations of the methods presented in these
papers. Specifically we will consider multiplicity dependent
efficiencies as well as a non-binomial 
efficiency distributions. We will discuss the most simple and
straightforward methods to implement those corrections. 
\end{abstract}
\maketitle

\section{Introduction}

Cumulants of conserved charges, such as the
baryon-number, are important observables in the search for a possible
phase structure in the QCD phase diagram
\cite{Stephanov:2008qz,Koch:2008ia}, and first measurements of the
net-proton and net-charge cumulants up to fourth order have been
carried out by the STAR collaboration
\cite{Xu:2014jsa,Adamczyk:2014fia,Adamczyk:2013dal}. As
pointed out in \cite{Bzdak:2012ab} finite detection efficiencies give rise to
fluctuations of the measured particle number and need to be corrected
for. These corrections can be included in a straightforward manner if the efficiency
follows a binomial distribution \cite{Bzdak:2012ab}. Appropriate
formulas for a phase space dependent (binomial) efficiency have also
been derived \cite{Bzdak:2013pha,Luo:2014rea,Kitazawa:2016awu}. These corrections
can be sizable as seen in the recent preliminary data by the STAR
collaboration \cite{Luo:2015ewa}. 

So far, however, all efficiency corrections have assumed that the
efficiency (in a given phase space bin) is constant for a given
centrality class, i.e. it does not depend on the multiplicity of particles under consideration. 
Furthermore, all the corrections have been carried
out assuming that the detection efficiency follows a binomial
distribution. 

In reality, the efficiency does depend on the multiplicity of
particles (see e.g. Fig. 1 in \cite{Luo:2015ewa}). Also it is not at
all obvious if a binomial distribution correctly describes the
detection probability. In the following we want to address both these
issues and discuss methods how to improve the efficiency corrections.

Let us start by defining more precisely what we mean by an
``efficiency distribution''. For simplicity, we will restrict
ourselves to one kind of particles, such as protons, and ignore
anti-particles. The extension to net-particles distributions is
straightforward following the discussion in
\cite{Bzdak:2012ab,Bzdak:2013pha}. Let us denote the number
distribution of the {\em produced} particles by $P(N)$ and that of the
{\em observed} particles by $p(n)$\footnote{Throughout this paper we
  will use lower case characters to refer to observed particles and
  upper case characters to refer to produced particles.}. 
Then the observed distribution is given by
\begin{align}
  p(n) = \sum_{N} B\left( n,N;\epsilon \right) P(N) , \label{eq:p_n}
\end{align}
where $B(n,N;\epsilon)$ denotes the probability to observe $n$ particles if $N$
particles are produced. The probability $B(n,N;\epsilon)$ depends on the detection efficiency
$\epsilon$. 
It is this probability $B(n,N;\epsilon)$ that we call
efficiency distribution.  
The detection efficiency, $\epsilon$, is given by the ratio of
mean number of {\em observed} particles, $\ave{n}$, over the mean
number of produced particles, $\ave{N}$, $ \epsilon = \ave{n} /
\ave{N}$. Obviously, $\epsilon$ by itself, does not define the entire
efficiency distribution. In practice however,  
$B$ is typically {\em assumed} to be a binomial distribution
\begin{align}
  B(n,N;\epsilon)=\frac{N!}{n! \left( N-n \right)!} \epsilon^{n}\left(
  1-\epsilon \right)^{N-n}, 
\label{eq:binomial}
\end{align}
in which case the knowledge $\epsilon$ is sufficient to characterize the distribution.
To which extent such an assumption is valid can only be verified by a
detailed simulation of a given detector system. 

Usually, the efficiency $\epsilon$ is assumed to be constant, i.e. independent of $N$.
In this case, factorial moments of the produced and observed particles
are simply related by \cite{Bzdak:2012ab}
\begin{equation}
f_{i}=\epsilon ^{i}F_{i},
\label{eq:fact_moments_relate}
\end{equation}%
where the factorial moments are defined by
\begin{align}
  F_{i}=\sum\nolimits_{N}P(N)\frac{N!}{\left( N-i\right) !},\quad
  f_{i}=\sum\nolimits_{n}p(n)\frac{n!}{\left( n-i\right) !}.
\label{eq:fact_moment_define} 
\end{align}
Given the above relation for the factorial moments, efficiency
corrections for the various cumulants are readily derived
\cite{Bzdak:2012ab}. However, in reality the efficiency may depend on
the multiplicity of particles under consideration. In addition, as
already mentioned, the
efficiency distribution may not be exactly binomial. In these cases, the above
simple formula (\ref{eq:fact_moments_relate}) will not hold and, as we will show in this paper, may
lead to wrong conclusions.  

Recent preliminary
results by the STAR collaboration \cite{Luo:2015ewa} at the lowest
available RHIC energies show that
the efficiency corrections play a crucial role in a proper interpretation of
data. Therefore, it is essential that the correct unfolding procedure
is applied. It is the purpose of this note to discuss various
corrections and modifications to the unfolding procedure, and we
should point out that we will only discuss the most simple and
straightforward unfolding methods and apply them to cumulants. 
There are other,  more refined, methods used to correct multiplicity
  distributions (see
  e.g. \cite{DAgostini:1994zf,Khachatryan:2010nk,Schmitt:2012kp}). 
 However, we are not aware that their suitability for the
  determination of higher order cumulants have so far been explored, and we
  hope that this note may motivate some work in applying these other
  methods to cumulant measurements. 

This paper is organized as follows: In the next section we will show
how the dependence of the efficiency on the number of particles
changes the results. After that we will discuss the effect of
non-binomial efficiency distributions by studying a few alternative
distributions. Then we will discuss the simplest unfolding procedure.
We will finish with a few comments and conclusions.

\section{Multiplicity dependent efficiency}
\label{sec:mult_dep}

In most experiments, the efficiency depends on the number of particles
in the detector. This is also the case for the  STAR experiment,
where the efficiency does
depend on the total number of charged particles, and thus may also
depend on the number of particles under consideration, $N$, such as
protons. While this does not
preclude the distribution $B$ from having the
binomial form, Eq.~\eqref{eq:binomial}, we now  have an efficiency $\epsilon(N)$ that
depends on the number of produced particles, $N$. Consequently the
relation between the factorial moments 
\begin{equation}
f_{i}=\sum\nolimits_{N}\epsilon ^{i}(N)P(N)\frac{N!}{\left( N-i\right) !},
\label{eq:fi_general} 
\end{equation}
is not as simple as in
Eq.~\eqref{eq:fact_moments_relate}. Furthermore, the unfolding
derived in \cite{Bzdak:2012ab} and used by the STAR collaboration
\cite{Adamczyk:2014fia} will not be possible anymore, even for a binomial
efficiency distribution $B$.

\begin{figure}[t!]
\begin{center}
\includegraphics[scale=0.4]{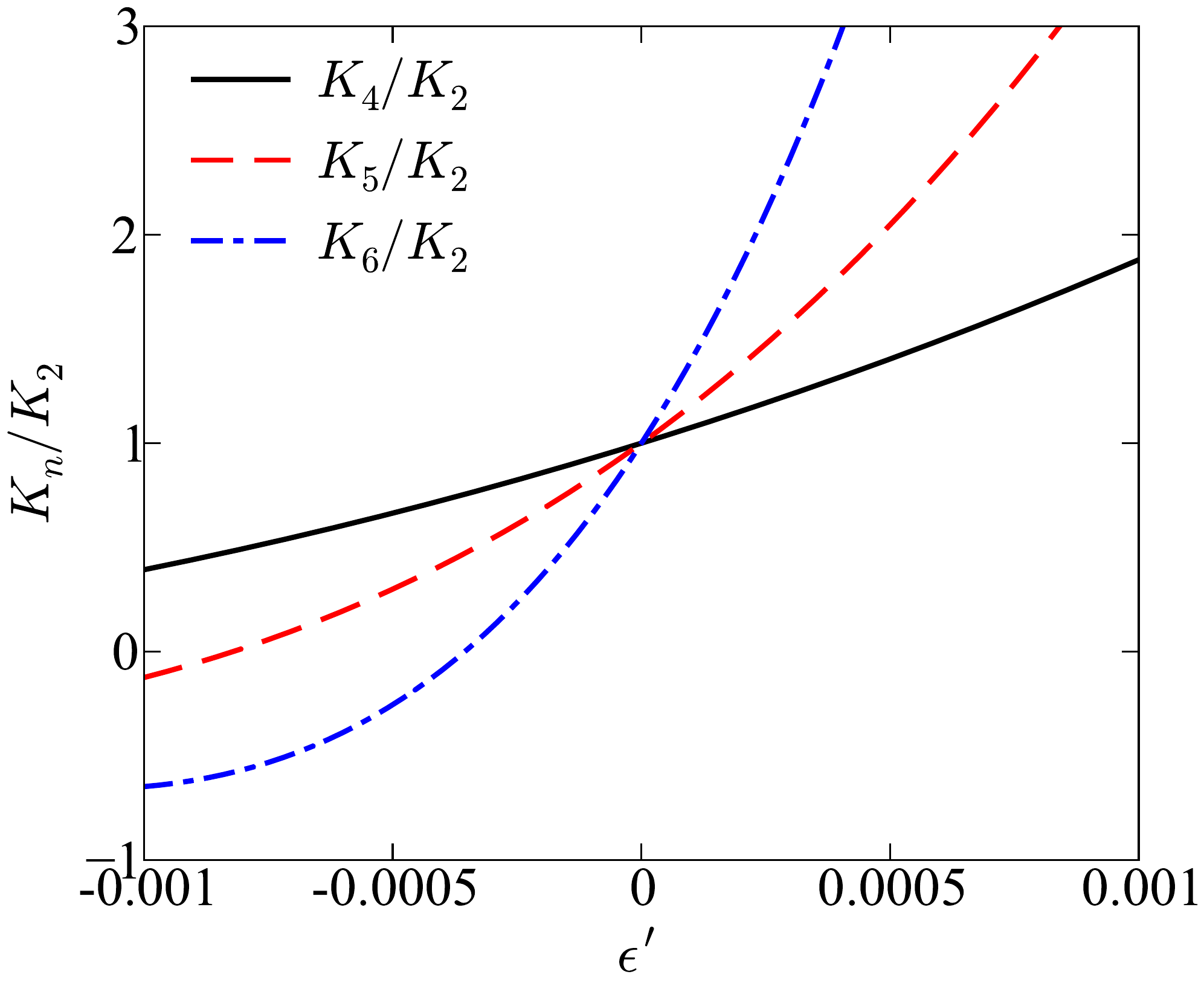}
\end{center}
\par
\vspace{-5mm}
\caption{$K_{n}/K_{2}$, $n=4$, $5$, $6$, as a function of $\protect\epsilon %
^{\prime }$ when corrected using average, multiplicity independent,
efficiency, $\protect\epsilon _{0}$, where in reality efficiency depends on
the number of produced protons, $\protect\epsilon (N)=\protect\epsilon _{0}+%
\protect\epsilon ^{\prime }(N-\left\langle N\right\rangle )$. In this
calculation $\protect\epsilon _{0}=0.65$ and $\left\langle N\right\rangle
=40 $ which are roughly the numbers for the STAR measurement at
low energies.}
\label{fig:1}
\end{figure}

In order to estimate the effect of a multiplicity dependent
efficiency, let us consider a simple example based on a Poisson
distribution for the produced particles and assume that the
efficiency depends linearly on the number of produced particles $N$,
\begin{eqnarray}
P(N) &=&\frac{\left\langle N\right\rangle ^{N}}{N!}e^{-\left\langle
N\right\rangle },  \notag \\
\epsilon (N) &=&\epsilon _{0}+\epsilon ^{\prime }(N-\left\langle
N\right\rangle ),  \label{eps-N}
\end{eqnarray}%
where $\epsilon_{0}$ is the average efficiency $\epsilon _{0}=\sum_{N}P(N)\epsilon
(N)$. In this case, the true cumulants ratios $K_{4,5,6}/K_{2}$ equal $1$. 
Using Eq. (\ref{eq:fi_general}) the factorial moments of the observed distribution are then
\begin{eqnarray}
f_{1} &=&\left\langle N\right\rangle (\epsilon _{0}+\epsilon ^{\prime }), 
\notag \\
f_{2} &=&\left\langle N\right\rangle ^{2}\left[ (\epsilon _{0}+2\epsilon
^{\prime })^{2}+\left\langle N\right\rangle (\epsilon ^{\prime })^{2}\right]
,  \notag \\
f_{3} &=&\left\langle N\right\rangle ^{3}\left[ (\epsilon _{0}+3\epsilon
^{\prime })^{3}+\left\langle N\right\rangle (\epsilon ^{\prime
})^{2}(3\epsilon _{0}+10\epsilon ^{\prime })\right] ,  \notag \\
f_{4} &=&\left\langle N\right\rangle ^{4}\left[ (\epsilon _{0}+4\epsilon
^{\prime })^{4}+\left\langle N\right\rangle (\epsilon ^{\prime
})^{2}(6\epsilon _{0}^{2}+52\epsilon _{0}\epsilon ^{\prime }+113(\epsilon
^{\prime })^{2}+3\left\langle N\right\rangle (\epsilon ^{\prime })^{2})%
\right] ,
\label{eq:first_moments}
\end{eqnarray}%
and more complicated formulas for $f_{5}$ and $f_{6}$. Now we can correct
using constant efficiency $F_{i}=f_{i}/\epsilon _{0}^{i}$, as described in
Ref. \cite{Bzdak:2012ab} and calculate all cumulants. The obtained results are presented
in Fig.~\ref{fig:1}, where we show $K_{4}/K_{2},$ $K_{5}/K_{2}$ and $%
K_{6}/K_{2}$ as a function of $\epsilon ^{\prime }$. Obviously for $\epsilon
^{\prime }=0$ our procedure is exact, $\epsilon (N)=\epsilon _{0}$, and we
obtain $K_{4,5,6}/K_{2}=1$. Interestingly even a very small $\epsilon
^{\prime }$ leads to substantial deviation from unity.

To put things
in perspective, for the STAR measurement at $\sqrt{s}=7.7 \GeV$,
$\epsilon' \simeq -0.1/250 \simeq -4 \times 10^{-4}$ so that the correction
for the ratio of $K_{4}/K_{2}$ is about 30\% and much larger for $K_{6}/K_{2}$.\footnote{%
We note that better results are obtained when we use an effective constant
efficiency $F_{i}=f_{i}/(\epsilon _{0}+i\epsilon ^{\prime })^{i}$, as
can be seen from Eq.~\eqref{eq:first_moments}.} 

While the 30\% correction for $K_{4}/K_{2}$ may not seem like much, we should keep in
mind that we have used a simple Poisson distribution to illustrate
things. In reality, especially if we are close to a critical
point, the true distribution will be far from Poisson and we need to be
able to unfold in a reliable way. As already mentioned, the analytic
methods described in \cite{Bzdak:2012ab} cannot be applied the moment
we have a multiplicity dependent efficiency. In
Section~\ref{sec:solve}, we will explore other means 
of unfolding the probability distribution.

\subsection{Non-binomial distributions}

Next let us explore what happens if the efficiency does not follow a
binomial distribution. To this end we will calculate the factorial
moments and subsequent cumulants for other non-binomial
distributions. Here we chose the hypergeometric, the beta-binomial
and the Gaussian distributions. 
The first two have limits
corresponding to the binomial distribution, allowing us to study the
deviations from binomial systematically. 
Our choice of non-binomial distributions is by no means motivated by
any possible detector design or effect, but simply to estimate the
effect of a possible non-binomial distribution.
Our strategy
is to calculate the factorial moments $f_{i}$ using Eq.~\eqref{eq:p_n} with
these non-binomial distributions $B(n,N)$ and unfold them using the
formula for the binomial distribution with constant efficiency,  
$F_{i}=f_{i}/\epsilon^{i}$. As ``input'' distribution for the {\em
  produced } particles we chose again a Poisson  distribution, 
with $\left\langle   N\right\rangle =40$. Therefore, the fact that we
unfold a non-binomial distribution using formulas based on the
binomial distribution will be reflected by
deviations from $K_{n}/K_{2}=1$. In order to isolate non-binomial effects
from issues related to multiplicity dependant efficiency, which we
have discussed previously,  we ensure that for the 
non-binomial $B(n,N)$ distributions the effective efficiencies given
by 
\begin{eqnarray}
\epsilon(N)=\frac{\ave{n}_{N}}{N}=\frac{1}{N}\sum_{n} n B\left( n,N
  \right) = \mathrm{constant}
\end{eqnarray}
do not depend on $N$.

\subsubsection{Hypergeometric distribution}
\begin{figure}[h]
\begin{center}
\includegraphics[scale=0.4]{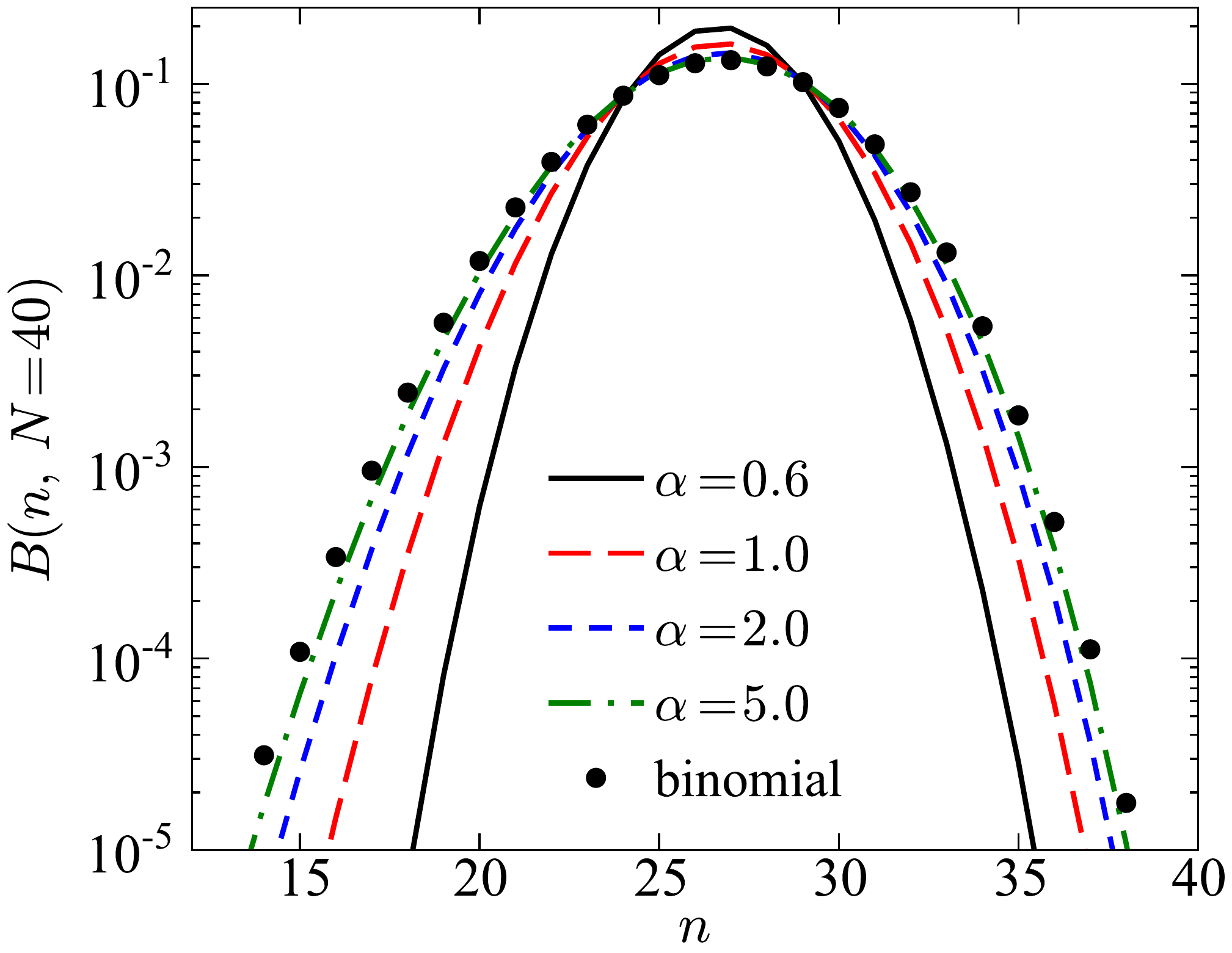} 
\end{center}
\par
\vspace{-5mm}
\caption{The hypergeometric distribution for different values of $\protect%
\alpha $ compared with the binomial distribution (black points). Here $N=40$
and $\protect\epsilon =2/3$.}
\label{fig:hyper}
\end{figure}

As the first example we consider the hypergeometric distribution. Suppose we
have an urn with $N_{w}$ white balls and $N_{b}$ black balls. For each
produced particle we sample a ball, if it is white we accept a particle and
if it is black we reject. In the case of the binomial distribution we return
balls to the urn and for the hypergeometric distribution balls are not
returned. In this case once we accept a particle (a white ball is removed
from the urn) the probability to accept the next one is a bit smaller. The
initial probability to accept a particle is given by $N_{w}/(N_{w}+N_{b})$.
The probability to accept $n$ particles at a given produced $N$ is given by (%
$N\leq N_{w}+N_{b}$) 
\begin{equation}
B(n,N)=\frac{1}{\binom{N_{w}+N_{b}}{N}}\binom{N_{w}}{n}\binom{N_{b}}{N-n},
\label{eq:hyper}
\end{equation}%
where we chose
\begin{equation}
N_{w}=2\alpha N,\quad N_{b}=\alpha N.
\end{equation}%
In this case 
\begin{equation}
\frac{\left\langle n\right\rangle _{N} }{N}=\frac{N_{w}}{N_{w}+N_{b}}=\frac{2}{3}
\end{equation}%
for each value of $N$, which corresponds to $\epsilon =2/3$ for the binomial
distribution. We note that in the limit of  $\alpha \rightarrow \infty$ the
hypergeometric distribution approaches a binomial.\footnote{For $N_{w}/N \rightarrow \infty$ 
and $N_{b}/N \rightarrow \infty$ the fact that balls are not returned to the urn is irrelevant.} 
In Fig. \ref{fig:hyper} we
show several curves for different values of $\alpha $ and fixed $N=40$. As
seen the hypergeometric distribution results in a narrower distribution than
binomial\footnote{By
  expressing $B(n,N)$, Eq.~(\ref{eq:hyper}), in terms of $\Gamma$
  functions, one is not restricted to integer values for $N_{b}$ and
  $N_{w}$ allowing to consider rather narrow distribution such as
  the example of $\alpha=0.6$ discussed here. }.

Finally we compute $p(n)$ using Eq. (\ref{eq:p_n}) and calculate the
factorial moments $f_{i}$. Next we correct them $F_{i}=f_{i}/\epsilon ^{i}$
and obtain the values presented in Tab. \ref{tab:hyper}.

\begin{table}[h]
\begin{center}
\begin{tabular}{|c|c|c|c|c|}
\hline
Hypergeometric & $\alpha =0.6$ & $\alpha =1.0$ & $\alpha =2.0$ & $\alpha =5.0
$ \\ \hline
$K_{3}/K_{2}$ & 1.16 & 1.12 & 1.07 & 1.03 \\ \hline
$K_{4}/K_{2}$ & 0.66 & 0.88 & 0.98 & 1.00 \\ \hline
$K_{5}/K_{2}$ & 2.19 & 1.68 & 1.23 & 1.05 \\ \hline
$K_{6}/K_{2}$ & -3.99 & -1.38 & 0.31 & 0.89 \\ \hline
\end{tabular}%
\end{center}
\par
\vspace{-5mm}
\caption{The obtained values of $K_{n}/K_{2}$ for the hypergeometric
distribution, using $F_{i}=f_{i}/\protect\epsilon ^{i}$ with $\protect%
\epsilon =2/3$, for different values of $\protect\alpha $ as presented in
Fig. \ref{fig:hyper}. }
\label{tab:hyper}
\end{table}

\subsubsection{Beta-binomial distribution}

The beta-binomial distribution is obtained from the binomial one when the
binomial success probability is random and follows the beta distribution.
Another interpretation (for positive integer $\alpha $ and $\beta $ being
the numbers of white and black balls, respectively) is similar to the
hypergeometric distribution however in this case once a white ball is drawn
two white balls are returned to the urn (and similar for black balls). The
resulting distribution of $n$ at a given $N$ is broader than binomial and is
given by 
\begin{equation}
B(n,N)=\binom{N}{n}\frac{B_{\text{eta}}(n+\alpha ,N-n+\beta )}{B_{\text{eta}%
}(\alpha ,\beta )}.
\end{equation}%
where
\begin{equation}
B_{\text{eta}}(x,y)= \int_{0}^{1}t^{x-1}\left( 1-t
\right)^{y-1}\,dt=\frac{\Gamma(x) \Gamma(y)}{\Gamma(x+y)}
\end{equation}
is the beta function or Euler integral of the first kind.
Taking%
\begin{equation}
\beta =\alpha \frac{1-\epsilon }{\epsilon },
\end{equation}%
we obtain%
\begin{equation}
\frac{\left\langle n\right\rangle _{N} }{N}=\epsilon ,
\end{equation}%
that is the efficiency does not depend on $N$. When $\alpha \rightarrow
+\infty $ the beta-binomial distribution goes into binomial. In Fig. (\ref%
{fig:beta}) we present four curves for $N=40,$ $\epsilon =0.7$ and different
values of $\alpha $.

Assuming the beta-binomial distribution we compute $p(n)$ using Eq. (\ref%
{eq:p_n}) with $P(N)$ given by Poisson and calculate the factorial moments $%
f_{i}$. Using $F_{i}=f_{i}/\epsilon ^{i}$ we obtain the values presented in
Tab. \ref{tab:beta}.

\begin{figure}[h]
\begin{center}
\includegraphics[scale=0.4]{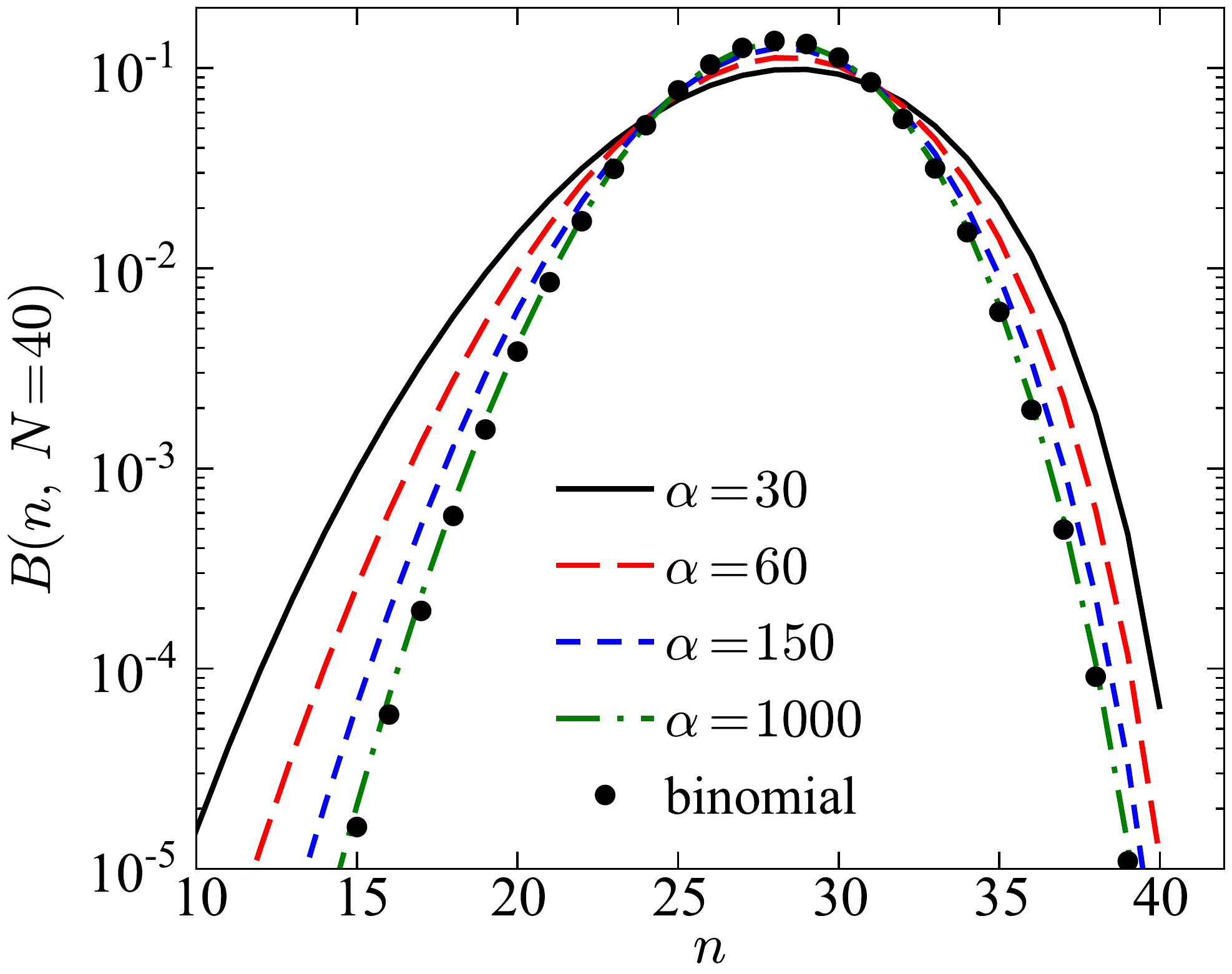}
\end{center}
\par
\vspace{-5mm}
\caption{The beta-binomial distribution for different values of $\protect%
\alpha $ compared with the binomial distribution (black points). Here $N=40$
and $\protect\epsilon =0.7$.}
\label{fig:beta}
\end{figure}

\begin{table}[h]
\begin{center}
\begin{tabular}{|c|c|c|c|c|}
\hline
Beta-binomial & $\alpha =30$ & $\alpha =60$ & $\alpha =150$ & $\alpha =1000$
\\ \hline
$K_{3}/K_{2}$ & 1.28 & 1.24 & 1.13 & 1.02 \\ \hline
$K_{4}/K_{2}$ & 0.82 & 1.45 & 1.35 & 1.07 \\ \hline
$K_{5}/K_{2}$ & -1.11 & 1.15 & 1.63 & 1.16 \\ \hline
$K_{6}/K_{2}$ & 5.71 & -0.44 & 1.80 & 1.32 \\ \hline
\end{tabular}%
\end{center}
\par
\vspace{-5mm}
\caption{The obtained values of $K_{n}/K_{2}$ for the beta-binomial
distribution, using $F_{i}=f_{i}/\protect\epsilon ^{i}$ with $\protect%
\epsilon =0.7$, for different values of $\protect\alpha $ as presented in
Fig. \ref{fig:beta}. }
\label{tab:beta}
\end{table}

\subsubsection{Gaussian distribution}

\begin{figure}[h]
\begin{center}
\includegraphics[scale=0.4]{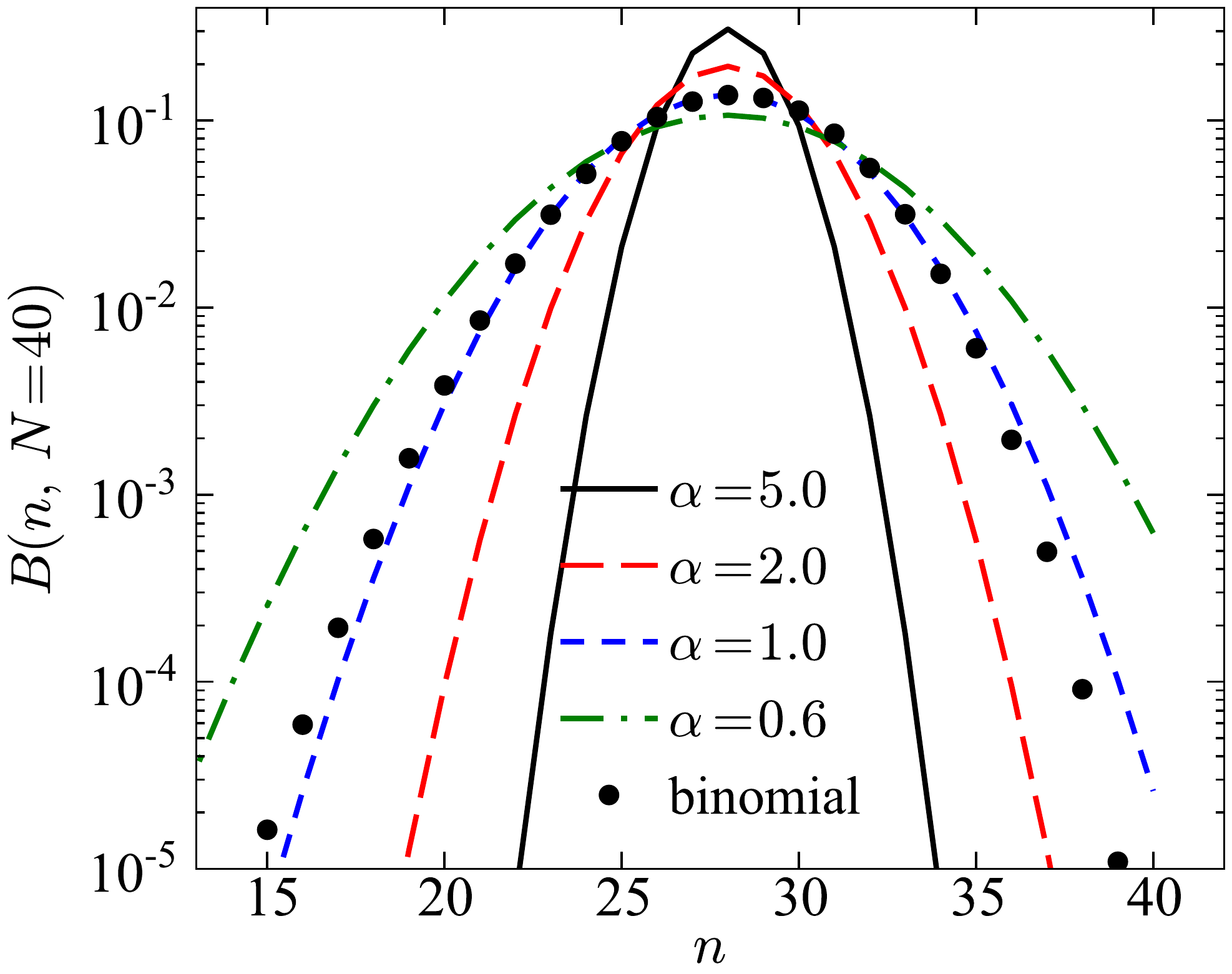}
\end{center}
\par
\vspace{-5mm}
\caption{The Gaussian distribution for different values of $\protect\alpha $
compared with the binomial distribution (black points). Here $N=40$ and $%
\protect\epsilon =0.7$.}
\label{fig:gauss}
\end{figure}

As the last example we consider the Gaussian distribution%
\begin{equation}
B(n,N)=\mathcal{N}(N,\epsilon ,\alpha )\exp \left( -\alpha \frac{%
(n-N\epsilon )^{2}}{2N\epsilon (1-\epsilon )}\right) \Theta\left( N-n \right),
\end{equation}%
where $\mathcal{N}$ is a normalization factor ensuring
$\sum_{n=0}^{N}B(n,N)=1$ and we enforce $B(n,N)=0$ for $n>N$. For
this distribution we approximately have (provided $\alpha $ is not too small)%
\begin{equation}
\frac{\left\langle n\right\rangle _{N} }{N}\simeq \epsilon ,
\end{equation}%
except for small values of $N$ (which is of no interest since we consider $%
\left\langle N\right\rangle =40$). In Fig. \ref{fig:gauss} we present four
curves for different values of $\alpha $ and in Tab. \ref{tab:gauss} we show
the corresponding values of cumulant ratios.

To summarize non-binomial distributions result in $K_{n}/K_{2}$ which
are different from one, as expected. However it is a somewhat
surprising and of course encouraging that for
small deviations from binomial the effect on $K_{4}/K_{2}$ is rather week,
especially if distributions are narrower than binomial. We note that
we also checked distributions for the produced particles other than
Poisson and found qualitatively similar effects.

\begin{table}[h]
\begin{center}
\begin{tabular}{|c|c|c|c|c|}
\hline
Gaussian & $\alpha =5.0$ & $\alpha =2.0$ & $\alpha =1.0$ & $\alpha =0.6$ \\ 
\hline
$K_{3}/K_{2}$ & 1.00 & 1.12 & 1.24 & 1.33 \\ \hline
$K_{4}/K_{2}$ & 0.54 & 0.93 & 1.58 & 2.22 \\ \hline
$K_{5}/K_{2}$ & 1.40 & 1.77 & 2.30 & 0.57 \\ \hline
$K_{6}/K_{2}$ & -1.97 & -0.46 & 3.31 & -14.3 \\ \hline
\end{tabular}%
\end{center}
\par
\vspace{-5mm}
\caption{The obtained values of $K_{n}/K_{2}$ for the Gaussian distribution,
using $F_{i}=f_{i}/\protect\epsilon ^{i}$ with $\protect\epsilon =0.7$, for
different values of $\protect\alpha $ as presented in Fig. \ref{fig:gauss}. }
\label{tab:gauss}
\end{table}

\section{Unfolding Methods}
\label{sec:solve}
Obviously the analytic formulas of Ref.~\cite{Bzdak:2012ab} are rather
limited and for the more general case of either multiplicity
dependent efficiency or non-binomial efficiency distributions, we need
different methods to unfold the measured distributions and/or cumulants. In this Section we demonstrate that the simplest unfolding method, based on solving triangular equations, result in correct cumulants even though the obtained multiplicity distribution, $P(N)$, is usually unphysical. 

Let us first see what happens when we try to unfold the entire
multiplicity distribution for the binomial efficiency distribution.

\subsection{Multiplicity distribution}

Our starting relation is 
\begin{equation}
p(n)=\sum_{N=n}^{\infty }P(N)\frac{N!}{n!(N-n)!}\epsilon ^{n}(1-\epsilon
)^{N-n},  \label{pP}
\end{equation}%
which can also be cast in matrix form
\begin{equation}
p(n) = B(n,N) P(N),  \label{pP-matrix}
\end{equation}%
where elements of $B$ are given by Eq. (\ref{eq:binomial}). $p$ is the measured 
distribution and $P$ is the true one. To make
analytical calculations we assume that $\epsilon $ does not depend on $N$.
Later on we show numerical calculation with $\epsilon $ depending on
$N$. So the problem of unfolding the multiplicity distribution is
equivalent to inverting the above equation. We note, that although we
will assume here that $B(n,N)$ is given by binomial our discussion is valid for other choices as well, as long as
$B(n,N)$ is not a singular matrix.

Suppose that in our experiment we measure $n$ from $n=0$ to $n=M$,
where $M$ is sufficiently large so that $P\left( N \right) \simeq 0$ for all
$N>M$. In this case the matrix gets finite and for example if 
$M=4$ we have%
\begin{equation}
\left( 
\begin{array}{c}
p(0) \\ 
p(1) \\ 
p(2) \\ 
p(3) \\ 
p(4)%
\end{array}%
\right) =\left( 
\begin{array}{ccccc}
1 & 1-\epsilon & (1-\epsilon )^{2} & (1-\epsilon )^{3} & (1-\epsilon )^{4}
\\ 
0 & \epsilon & 2\epsilon (1-\epsilon ) & 3\epsilon (1-\epsilon )^{2} & 
4\epsilon (1-\epsilon )^{3} \\ 
0 & 0 & \epsilon ^{2} & 3\epsilon ^{2}(1-\epsilon ) & 6\epsilon
^{2}(1-\epsilon )^{2} \\ 
0 & 0 & 0 & \epsilon ^{3} & 4\epsilon ^{3}(1-\epsilon ) \\ 
0 & 0 & 0 & 0 & \epsilon ^{4}%
\end{array}%
\right) \left( 
\begin{array}{c}
P(0) \\ 
P(1) \\ 
P(2) \\ 
P(3) \\ 
P(4)%
\end{array}%
\right) .
\end{equation}

Our goal is to solve equation (\ref{pP-matrix}) and obtain $P(N)$. One
immediate problem is that the matrix $B$ is practically singular in
realistic situations. Indeed, the determinant of the triangular matrix $B$
is given by a product of its diagonal elements. We obtain%
\begin{equation}
\det (B)=\prod\nolimits_{i=0}^{i=M}B(i,i)=\prod\nolimits_{i=0}^{i=M}\epsilon
^{i}=\epsilon ^{0+1+...+M-1+M}=\epsilon ^{M(M+1)/2}.
\end{equation}%
For example for $\epsilon =0.7$ and $M=100$ we obtain $\det (B)\sim
10^{-782} $, which is zero for all practical purposes. Consequently solving
Eq. (\ref{pP-matrix}) usually leads to unphysical $P(N)$. However we will
show later that even though $P(N)$ is usually unphysical the obtained
cumulants are correct.

In the case when $B$ is given by a binomial distribution the inverse
relation can be given analytically, 
\begin{equation}
P(N)=\sum_{n=N}^{\infty }p(n)\frac{n!}{N!(n-N)!}\frac{1}{\epsilon ^{n}}%
(-1+\epsilon )^{n-N},  \label{Pp}
\end{equation}
or in other words, the inverse of the binomial matrix is given by
\begin{align}
  B^{-1}(N,n) = \frac{n!}{N!(n-N)!}\frac{1}{\epsilon ^{n}}
(-1+\epsilon )^{n-N},
\end{align}
so that\footnote{We note that in practical applications inverting a pseudo-singular matrix $B$ is not advised. Instead, equations should be solved directly taking advantage of the fact that $B$ is triangular (by definition $N \ge n$).}
\begin{equation}
P(N)= B^{-1}(N,n) p(n),
\label{eq:P_from_p}
\end{equation}
or, explicitly, the first few terms,
\begin{equation}
\left( 
\begin{array}{c}
P(0) \\ 
P(1) \\ 
P(2) \\ 
P(3) \\ 
P(4)%
\end{array}%
\right) =\left( 
\begin{array}{ccccc}
 1 & \frac{\epsilon -1}{\epsilon } & \frac{(\epsilon -1)^2}{\epsilon ^2} & \frac{(\epsilon -1)^3}{\epsilon ^3} & \frac{(\epsilon -1)^4}{\epsilon
   ^4} \\
 0 & \frac{1}{\epsilon } & \frac{2 (\epsilon -1)}{\epsilon ^2} & \frac{3 (\epsilon -1)^2}{\epsilon ^3} & \frac{4 (\epsilon -1)^3}{\epsilon ^4} \\
 0 & 0 & \frac{1}{\epsilon ^2} & \frac{3 (\epsilon -1)}{\epsilon ^3} & \frac{6 (\epsilon -1)^2}{\epsilon ^4} \\
 0 & 0 & 0 & \frac{1}{\epsilon ^3} & \frac{4 (\epsilon -1)}{\epsilon ^4} \\
 0 & 0 & 0 & 0 & \frac{1}{\epsilon ^4} \\
\end{array}
\right)
 \left( 
\begin{array}{c}
p(0) \\ 
p(1) \\ 
p(2) \\ 
p(3) \\ 
p(4)%
\end{array}%
\right) .
\end{equation}
 
As seen from Eq. (\ref{Pp}), $P(N)$ is prone to large errors since we add
many terms of alternating sign. This is the main reason why $P(N)$ is usually
unphysical. This is especially problematic since $p(n)$ will only be
known within statistical uncertainties. However, as we will argue
below the resulting cumulants are usually correct (within statistical errors) even if $P(N)$ is unphysical.

\subsection{Cumulants}

\begin{figure}[t]
\begin{center}
\includegraphics[scale=0.35]{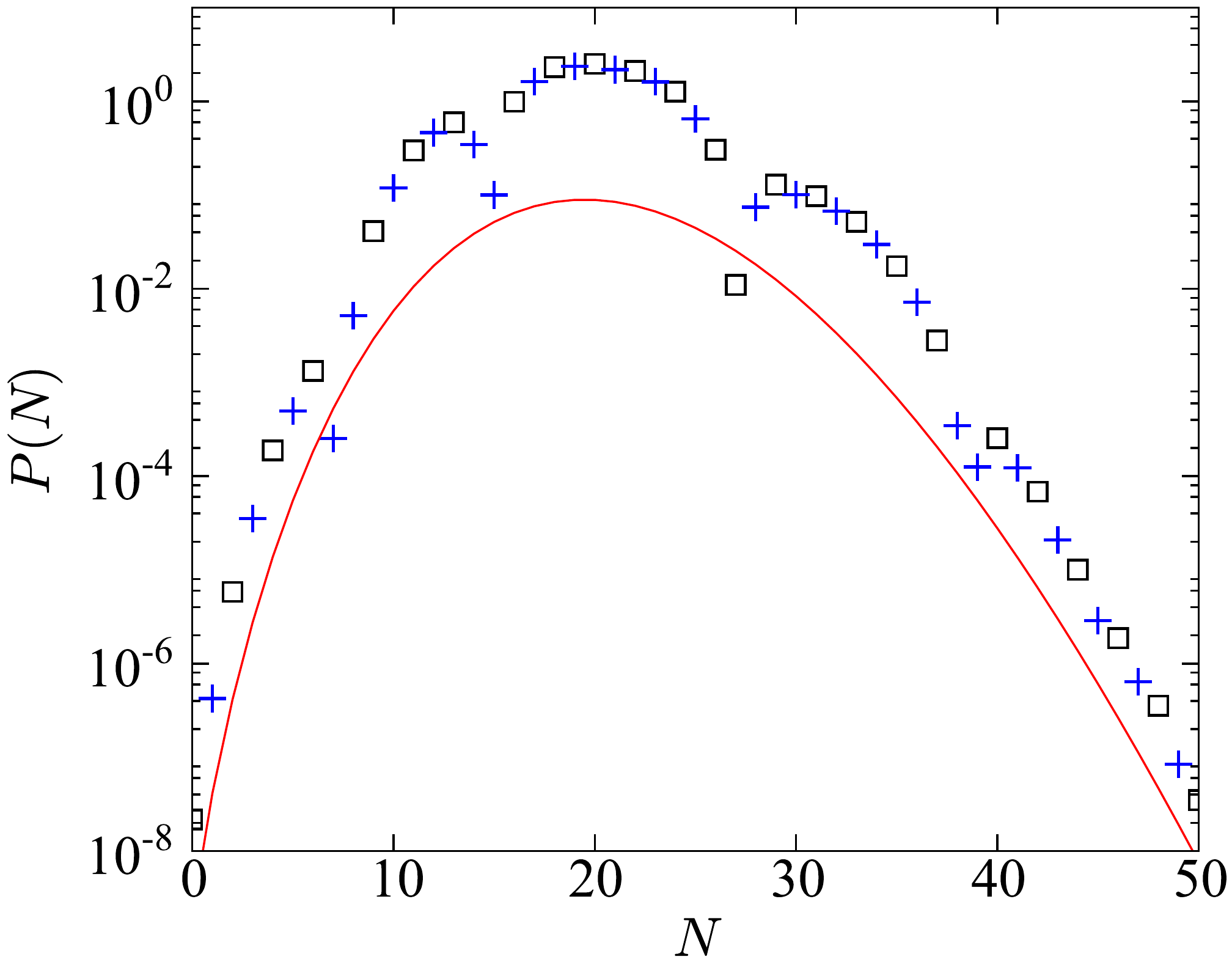}
\end{center}
\par
\vspace{-5mm}
\caption{The calculated $P(N)$ using the exact Eq. (\ref{Pp}) and the exact $p(n)$
given by Eq. (\ref{pP}) multiplied by $(1+\delta_{n})$, where $\delta_{n}$ is a random
number very close to zero. The solid red line represents the Poisson distribution, which is our
input. The blue crosses represent negative $P(N)$ and are shown as $%
-P(N) $. Positive $P(N)$ are shown as black open squares.
As discussed in the main text this unphysical $P(N)$ results, with a very good accuracy, in correct cumulants.}
\label{fig:2}
\end{figure}

As already pointed out, the sum in Eq.~(\ref{Pp}) has subsequent positive and
negative terms and there is a delicate cancellation between them. As a
consequence even for negligible ``noise'' on $p(n)$  
we get an incorrect $P(N)$, unless $\epsilon$ is very close to unity,
$\epsilon \simeq 1$. 
This is demonstrated in Fig.~\ref{fig:2}, where we have started out
with a Poisson distribution with $\langle N\rangle =20$ for the
produced particles. We then calculate the distribution of observed
particles, $p(n)$ using Eq.~(\ref{pP}) for $\epsilon =0.6$. We
introduce the ``noise'' by replacing $p(n)$ with $p(n)(1+\delta_{n})$,
where $\delta_{n}$ is sampled from the Gaussian distribution, $%
e^{-\delta_{n}^{2}/2\sigma ^{2}}$ with $\sigma =10^{-5}$. Finally we use Eq.~(\ref%
{Pp}) to obtain $P(N)$. Our result is presented in Fig.~\ref{fig:2}. 
The solid red line represents Poisson, which is our
input. The blue crosses represent negative $P(N)$ and are shown as $%
-P(N) $. Positive $P(N)$ are shown as black open squares.

However this problem vanishes once we sum over $N$. Indeed
the factorial moment $F_{i}$ is given by 
\begin{eqnarray}
F_{i} &\equiv &\sum_{N=i}^{\infty }P(N)\frac{N!}{(N-i)!}  \notag \\
&=&\sum_{N=i}^{\infty }\sum_{x=N}^{\infty }p(x)(1+\delta_{x})\frac{x!}{N!(x-N)!}%
\frac{1}{\epsilon ^{x}}(-1+\epsilon )^{x-N}\frac{N!}{(N-i)!}  \notag \\
&=&\sum_{x=i}^{\infty }p(x)(1+\delta_{x})\frac{1}{\epsilon ^{i}}\frac{x!}{(x-i)!}%
\simeq \frac{1}{\epsilon ^{i}}f_{i},
\end{eqnarray}%
and the noise $\delta_{n}$ does not affect the result in a significant
way since we add only positive numbers, as $|\delta_{n}|\ll1$. 

Consequently, while the extracted multiplicity distribution  $P(N)$ is
rather erratic, and clearly
unphysical, the  calculated cumulant ratio, $K_{4}/K_{2}$ equals $1$
with good accuracy.\footnote{%
When we use the analytic equation (\ref{pP}) for $p(n)$, namely $\delta_{n}=0$, and sum Eq. (\ref%
{Pp}) numerically we still obtain an unphysical multiplicity
distribution $P(N)$ for sufficiently small $%
\epsilon $. This is because we add very large numbers with opposite signs. In this case
the calculated cumulants are incorrect. This situation happens for $%
\langle N\rangle =20$ and $\epsilon <0.2$ or for $\langle
N\rangle =40$ and $\epsilon <0.6$, and is due to insufficient
numerical accuracy necessary to handle the $1/\epsilon^{n}$ terms. }

\subsection{Cumulants with multiplicity dependent efficiency}

In this Section we test the previously discussed method by using the multiplicity dependent
efficiency given by Eq. (\ref{eps-N}). We sample the  produced
particles from a Poisson distribution $P(N)$ and we parameterize the
efficiency by $\epsilon (N)=\epsilon _{0}+\epsilon ^{\prime
}(N-\langle N\rangle )$. We use the following parameters for our
calculation:  $\left\langle
N\right\rangle =40$, $\epsilon _{0}=0.7$, $\epsilon ^{\prime }=-0.0005$ and 
$\left\langle N\right\rangle =20$, $\epsilon _{0}=0.6$, $\epsilon ^{\prime
}=-0.001$.  

\begin{figure}[h]
\begin{center}
\includegraphics[scale=0.4]{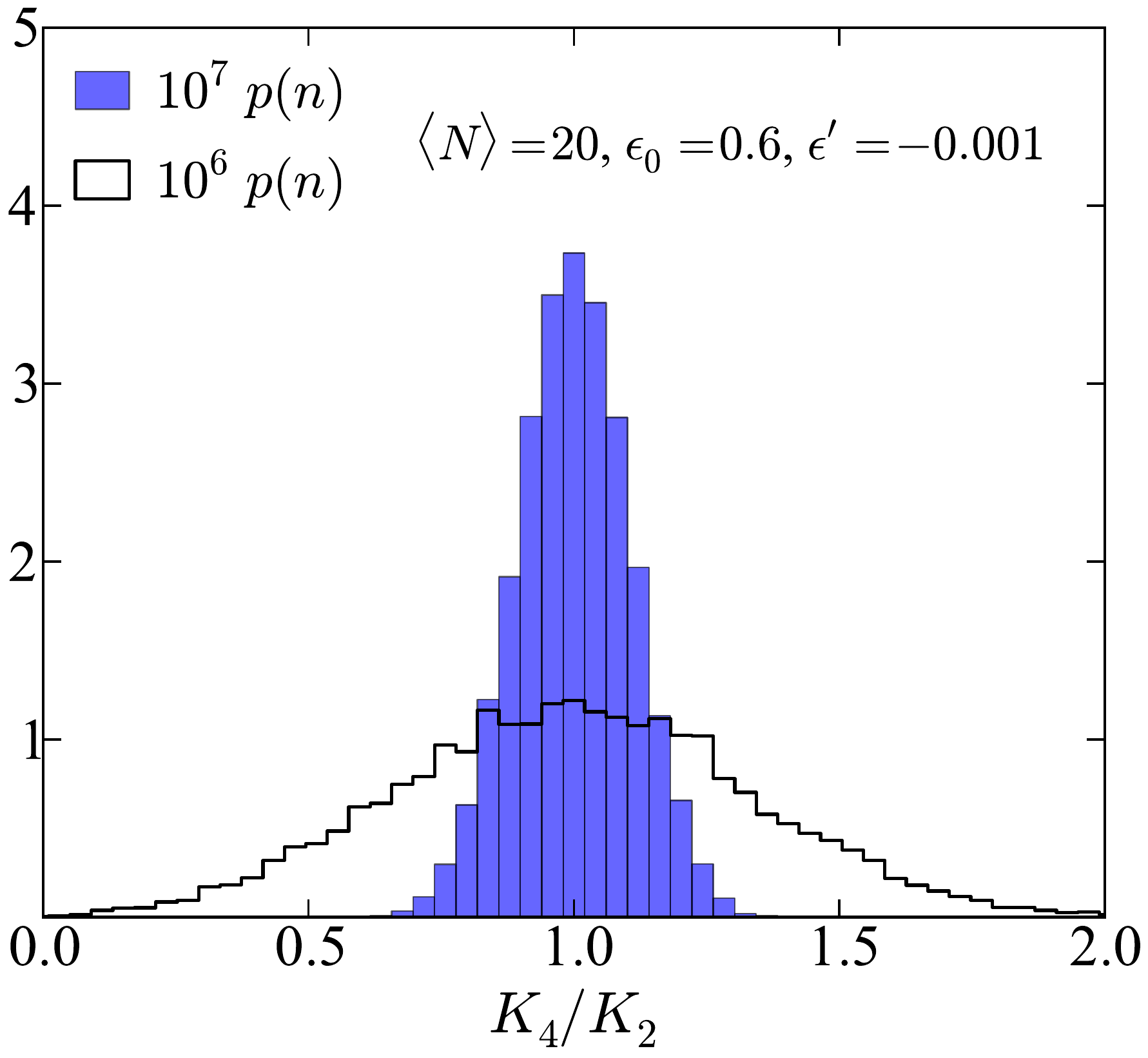}
\includegraphics[scale=0.4]{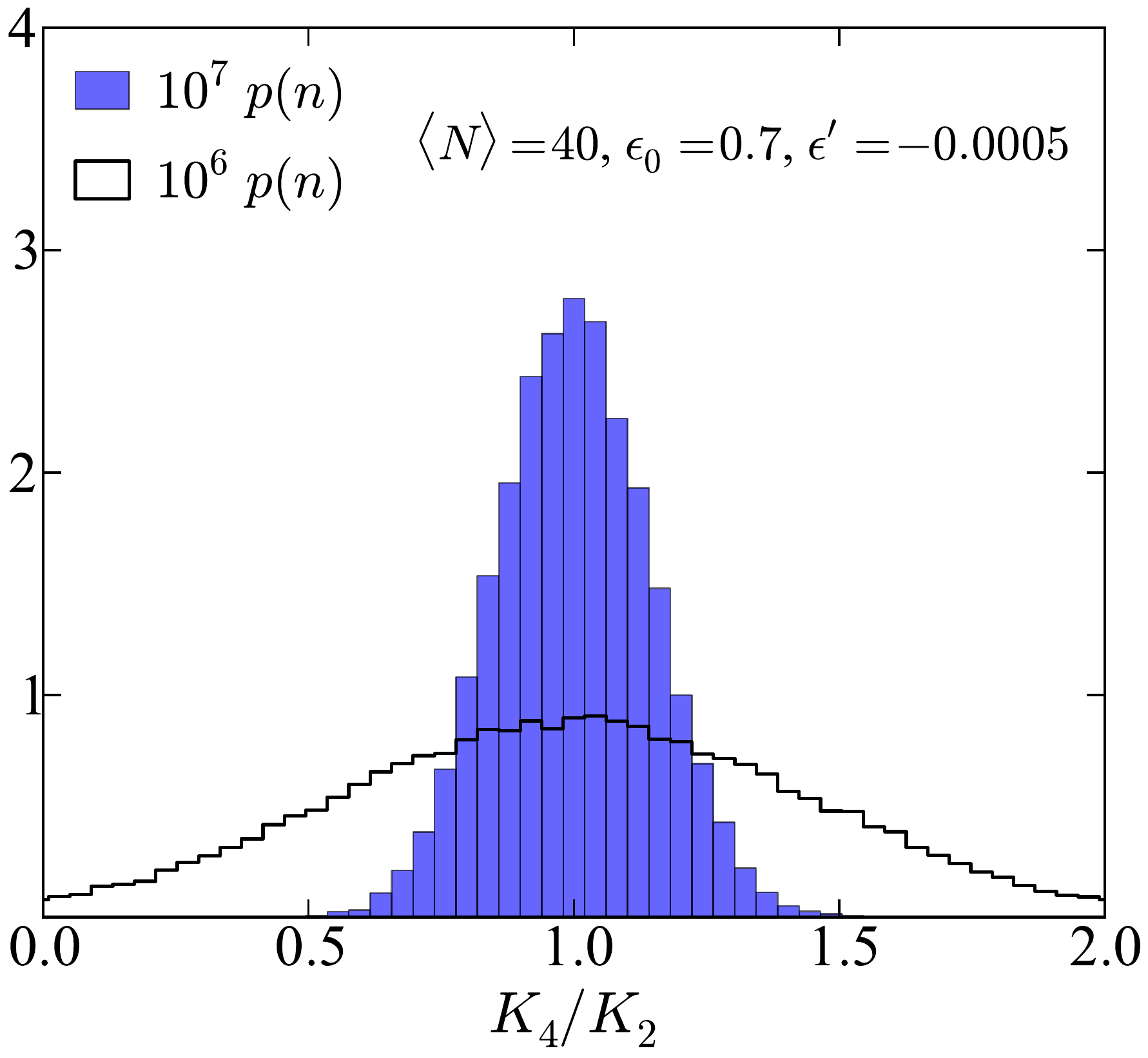}
\end{center}
\par
\vspace{-5mm}
\caption{Histogram (normalized to unity) of $K_{4}/K_{2}$ in the case of sampled $p(n)$ ($10^{7}$
events) and analytical matrix given by Eq. (\ref{B-N}) with $\epsilon (N)=\epsilon _{0}+\epsilon ^{\prime
}(N-\langle N\rangle )$. For comparison we
also show results with $10^{6}$ events.}
\label{fig:4}
\end{figure}

We first sample $N$ from the Poisson distribution. Next for each of
these $N$ particles we
decide whether it is detected or not with binomial probability $\epsilon (N)=\epsilon
_{0}+\epsilon ^{\prime }(N-\left\langle N\right\rangle )$. We run $10^{7}$
events which allows to calculate the measured $p(n)$. Our efficiency matrix
is given by%
\begin{equation}
B(n,N)=\frac{N!}{n!(N-n)!}\epsilon (N)^{n}\left[ 1-\epsilon (N)\right]
^{N-n}.  \label{B-N}
\end{equation}%
Next we solve Eq. (\ref{pP-matrix}) for the true distribution
$P(N)$. Here we take advantage of the fact that the matrix $B$ is triangular, which allows for straightforward backward substitution. We note there is no need to invert $B$, which is not advised for pseudo-singular matrices.\footnote{We checked that inverting $B$ and using Eq. (\ref{eq:P_from_p}) leads to unphysical cumulants.}
Finally we calculate the true $K_{4}/K_{2}$. We repeat this exercise
a few thousand times and plot histogram of the resulting cumulant ratios $K_{4}/K_{2}$. This is
presented in Fig.~\ref{fig:4} 

As seen from Fig.~\ref{fig:4} the cumulant ratios are centered around $1$ with
rather small statistical spread (for $10^{7}$ events in each experiment).
For comparison we also show calculations with $10^{6}$ events, which
obviously results in larger statistical error.

\begin{figure}[t]
\begin{center}
\includegraphics[scale=0.4]{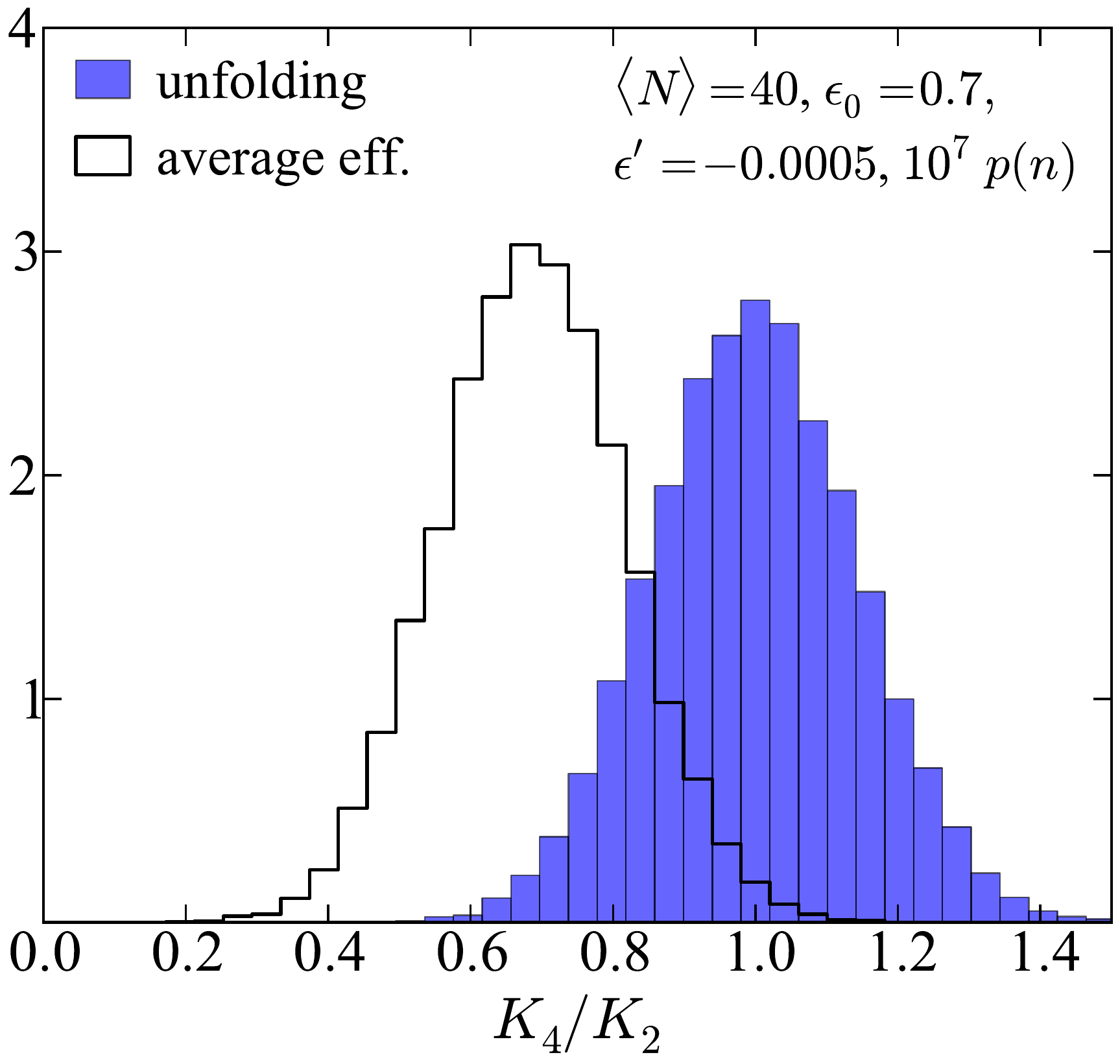}
\end{center}
\par
\vspace{-5mm}
\caption{Histogram (normalized to unity) of $K_{4}/K_{2}$ calculated using the sampled $p(n)$ and
(i) by solving Eq. (\ref{pP-matrix}) and (ii) by correcting factorial
moments using average constant efficiency. Both methods result in comparable
statistical errors.}
\label{fig:5}
\end{figure}

One could worry that this method introduces larger errors than the
method with constant efficiency proposed in Ref. \cite{Bzdak:2012ab}. We checked this explicitly and this is not
the case. We took the sampled $p(n)$, calculated $f_{i}$ and corrected them
using a constant average efficiency, $\epsilon _{0}$, so that $F_{i} =
f_{i}/ \epsilon_{0}^{i} $. The resulting cumulant ratios $K_{4}/K_{2}$
are shown in Fig. \ref{fig:5} as a black solid line. 
As discussed in Section 1, $K_{4}/K_{2}$ comes out too small (in agreement
with Fig. \ref{fig:1}). However the statistical error is comparable to the method
with solving triangular Eq. (\ref{pP-matrix}). 

The obvious advantage of solving Eq. (\ref{pP-matrix})
is that  we now obtain the correct cumulant ratio
$K_{4}/K_{2}$. Therefore, as long as we know the efficiency matrix and
if it is not singular, we can determine the cumulants of the original
distribution of produced particles, $P(N)$ within purely statistical uncertainties.

In principle one could simulate the efficiency matrix, $B(n,N)$, by careful analysis of
a given detector, for example using GEANT.
In this case we would run many
simulations and for each true $N$ we determine the measured probability to
observe $n$ particles resulting in the efficiency matrix $B(n,N)$,
which should reflect all detector effects (to the extent that they are
properly simulated in GEANT).
Unfortunately this method most likely results in the matrix $B$ being
mathematically singular. The problem is that with growing $N$ it is
getting very unlikely to observe $n=N$ (unless $\epsilon$ is very
close to $1$) and we get zeros for some diagonal elements. For a
triangular matrix it means that its determinant vanishes and that it
is therefore singular. One could cut the matrix so that it is
non-singular, however this results in too small
matrix.\footnote{Unless we can simulate our detector with very large, currently impossible, number of events. For example if $B$ is binomial with $\epsilon=0.7$, the diagonal elements are $B(n=N,N)=\epsilon^{N}$. For instance it gives roughly $10^{-10}$ for $N=65$.}

Let us briefly discuss two ways to overcome this difficulty. 
First, one could try to fit the simulated matrix $B(n,N)$ with some
function, for example a binomial distribution with $\epsilon$
depending on $N$. Using the analytical form for $B$ we obtain the full
matrix and we can successfully extract true cumulants, as discussed in
the previous Section. Of course this method is model-dependent and
relies on a correct extrapolation of $B$ to higher values of $N$. 

Another way is to keep the simulated (incomplete) $B(n,N)$ as it is and
solve the matrix equation using the singular value decomposition
(Moore--Penrose pseudoinverse). This method is designed to obtain a
solution from an under-determined set of equations (less equations than
unknowns, which is our case). We checked empirically that this method
reproduces a distribution of cumulant ratios which is also centered at
$K_{4}/K_{2}=1$, however, with very long tails, so that the width is
not only due to finite statistics.   

Obviously one may consider alternative methods, such as the one
already employed to extract the charged particle multiplicity
distribution, see e.g. \cite{Khachatryan:2010nk}, namely a Bayesian unfolding method
\cite{DAgostini:1994zf}, or other refined unfolding methods
\cite{Schmitt:2012kp}.  
However, we  are not aware of any work where this
method has been applied to the determination of higher order cumulants
and it would be worthwhile to assess its suitability.

\section{Conclusions }
\begin{itemize}
\item We stress that it is very unlikely that a 
  binomial distribution with constant efficiency is a correct model
  for the efficiency distribution. To the very least one likely has to
  take into account a multiplicity dependent efficiency. Given the
  substantial uncertainties demonstrated in this note, it appears
  mandatory that each experiment wishing to measure higher order
  cumulants needs to extract the full efficiency matrix, specific to this
  experiment. To which extent this can be done reliably, remains to be
  seen. Maybe the only solution for a credible measurement of higher
  order cumulants is to design a dedicated experiment which has an
  efficiency very close to $1$.  

\item As already stated in the introduction, in this note we studied
  the most straightforward method for the determination of the
  cumulants of the true distribution. However, we are aware that there 
  are more sophisticated unfolding methods,
  e.g. those applied in high-energy physics to various problems 
  (see
  e.g.  \cite{DAgostini:1994zf,Khachatryan:2010nk,Schmitt:2012kp}). To our
  knowledge these have not yet been applied to the determination of higher
  order cumulants, however,  and it would be worthwhile to study their
  suitability to do so.
  
\end{itemize}

\section*{Acknowledgments}
We thank A. Kalweit, J. Thaeder for useful discussions. We also
thank the HIC for FAIR and  ExtreMe Matter Institute (EMMI) for
support to attend two workshops where this work was initiated. 
AB was supported by the Ministry of Science and Higher Education (MNiSW), by funding from the Foundation for Polish Science, and by the National Science Centre (Narodowe Centrum Nauki), Grant No. DEC-2014/15/B/ ST2/00175, and in part by DEC-2013/09/B/ST2/00497. 
VK was supported by the Director, 
Office of Science, Office of High Energy and Nuclear Physics, 
Division of Nuclear Physics, and by the Office of Basic Energy
Sciences, Division of Nuclear Sciences, of the U.S. Department of Energy 
under Contract No. DE-AC03-76SF00098 and DOE Contract No. DE-AC
02-98CH10886.

%


\end{document}